# Commensurability effects induced by a periodic array of nanoscale anti-dots in Nb superconductor

A.A.Zhukov[a,][*], E.T.Filby[a], P. A. J. de Groot[a], V.V.Metlushko[b] and B.Ilic[c]

[a] Department of Physics and Astronomy, University of Southampton, SO17 1BJ, UK,

[b] Department of Electrical and Computer Engineering, University of Illinois at Chicago, Chicago, IL 60607-0024,

[b] Cornell Nanofabrication Facility and School of Applied and Engineering Physics, Cornell, University, Ithaca, NY 14853

**Abstract**

We study the interactions of the vortex lattice with a periodic square array of holes in a superconducting Nb film using magnetic and resistive measurements. Three different temperature regions have been observed. They are governed by interplay between vortex-vortex interactions and pinning by holes. At low temperatures flux jumps suppress the commensurability anomalies. In all measurements a peak has been observed close to commensurate states 5-7. The Cole-Cole plot reveals significant changes in the flux penetration mechanism at this point.

# 1. Introduction

The interaction between a periodic elastic medium and a periodic pinning potential represents the behavior of many important physical systems. Besides the case of the vortex lattice interacting with a periodic artificial pinning array in type II superconductors, it includes the behavior of vortices in layered superconductors with intrinsic pinning [1], charge [2] and spin [3] density waves, mass density waves in superionic conductors [4], polarization density waves in incommensurate ferroelectrics [1], absorbed atomic layers on periodical substrate [5], magnetic bubble arrays [6] and the magnetically induced Wigner crystal in a two dimensional electron gas [7]. The easy control of vortex separation in superconductors by simply varying the external field and the large diversity in possible interaction manifolds and in intrinsic parameters makes the vortex system one of the most versatile experimental testing fields.

Modern technologies allow us to prepare samples with an efficient periodic system of pinning centres. Among them much interest is generated by arrays of holes with submicron length scales. Interaction of the vortex system with such a regular array of anti-dots results in a wide range of commensurability effects observed for integer or rational number of flux quanta per pinning site [8-11]. Many interesting effects have been found, including the existence of multiquanta vortices on holes [9, 10], vortices taking interstitial positions [9, 12], Shapiro steps [13], field polarity dependent flux pinning [14], etc. Individual and multi-vortex pinning was considered in numerical simulations [15, 16]. However, several aspects related to the commensurate-incommensurate transition and transformation of multi-quanta into interstitial vortices remains to be clarified.

**2. Sample preparation and experimental techniques**

An anti-dot structure was prepared in a Nb superconducting film with a thickness of 100nm using laser interferometric lithography [17, 18]. This method allows us to approach an ideal periodicity on the scale of the whole sample. The high quality of the arrays prepared in this manner was evident from the large number of commensurate states observed [18, 19]. In the sample used in our investigations a periodic square lattice of holes with a diameter D of (350±50)nm and a period d of 1100 nm was prepared. The holes are penetrating through the full film thickness. Magnetic measurements have been performed using a Hall probe (sensitive area 50μm×50μm) magnetometer [20]. The ac-susceptibility has been measured from the Hall probe response to ac-magnetic field with a frequency of 315Hz and amplitude in the range 1μT - 1mT. The resistance was measured by a standard four probe technique using a square modulated current with a frequency of 68 Hz.

## 3. Experimental results and discussion

In our experiment we placed the sample with contacts above the Hall probe and simultaneously measured resistive voltage and Hall probe response to the applied ac-field $H_{ac}$. As can be seen from Fig.1 the vanishing point of the resistance is followed by a steep increase in the diamagnetic shielding. The onset of the superconducting transition was found to be $T_c$ = 7.88 K, and the resistive transition width between the levels of 10-90% is 0.09 K. The smaller value of $T_c$ in comparison with bulk Nb is typical for thin films [19, 21] and probably originates from a background oxygen contamination. A similar superconducting transition was found in plain film. The onset point for ac-susceptibility is 7.60K and the middle point corresponds to 7.43K. The ac-susceptibility transition width between the levels of 10-90% is 0.10 K. We did not find any interference between the two current systems. Neither switching off the ac-modulation or electric current affects the other transition. From the measurements of $T_c$ (middle point of ac-response) we find the $B_{c2}$ line. From the linear slope of $dB_{c2}/dT$ near $T_c$ we find that $B_{c2} = 0.7 T_c dB_{c2}/dT = 2.5T$. This gives us a coherence length $\xi(0) = 11.2$nm. Using parameters for very clean Nb $\xi_{BCS} = 40.8$nm and $\lambda_{BCS} = 47$nm [22] we can find the mean free path 4.3nm and $\lambda(0) = 90$nm. According to Ref.23 the maximum possible number of vortices per hole may be equal to $n_s = D / 4\xi(0) \approx 8\pm1$.

From previous results we see that magnetic and resistive measurements probe different temperature intervals. The presence of a periodical array of holes produces clear matching anomalies at $H_n = n\Phi_o/S$ with n = 0, ±1, ±2, …, $\Phi_o$ is the flux quantum and $S = d^2$ ( d is the period of the square structure)., for all different measurements. As can be seen from Fig.2 flux jumps dominate magnetization behaviour below 7.4K at small fields. With decreasing temperature the region of instability expands and below 6K wipes out all matching anomalies. For fields outside the instability region, the magnetization demonstrates step-like behaviour at matching points. This can be explained by the linear decrease of pinning force with number of vortices pinned on a hole $f_p \propto 1 - n/n_s$ [23]. In the single vortex pinning regime the dissipation is determined by the weakest pinning centres and the critical current is expressed by $j_c = f_p / \Phi_o$. Of course the full picture is rather complicated. In particular at very low temperatures a multi-terrace critical state [24] can be realised. In this case within the Bean flux profile the density of vortices changes by more than one flux quantum. However, in our sample the self-field value is smaller than 20Oe even at 6K. Therefore, this state should not be realised. Furthermore it is worth to mention that for increasing magnetic field, the commensurability anomalies remain weaker than for the decreasing branch of the magnetic field. This behaviour suggests the influence of the surface or edge barrier leading to more disorder in the vortex lattice for increasing magnetic field. We should also mention that all commensurability anomalies have a similar shape except for the steps with n=5-7. They are overlaid by a broad peak close to ~110Oe.

In another type of magnetic measurements using ac-susceptibility we can access slightly higher temperatures and separate shielding, reflected by the in-phase component $\chi'$, from dissipation measured by the out-of-phase component $\chi''$. In Fig.3 the real component $\chi'$ is presented for different temperatures. We should mention that the curves essentially coincide for increasing and decreasing magnetic field. Obviously in both cases the oscillatory vortex movement is essentially the same. In addition this ac-shaking may lead to improved order in the vortex system. Here we clearly see that with increasing temperature and magnetic field the step anomalies at matching fields start to transform into peaks. This has been first observed in Pb/Ge multilayers and related to a "superconducting networks regime" [24]. In other words this is regime of weak pinning when inter-vortex interaction becomes dominant. When $\xi$ exceeds the hole size D the pinning force starts to drop fast and finally leads to a thermally-activated depinning. In particular, for a columnar defect the pinning energy $V_p \approx \varepsilon_o (D/2\xi)^2$ with $\varepsilon_o = (\Phi_o /4\pi\lambda)^2$ [25], becomes much smaller than elastic energy $V_{el} \approx \varepsilon_o \log(\lambda/\xi)$ [26] per unit length of vortex line. In the case where elastic interactions become important the resulting pinning force deviates strongly from the single vortex pinning value and should reach a maximum for the commensurate state. Competition between the pinning forces of the periodical array and inter-vortex interaction of the elastic vortex system results into a locked state near commensurate fields [27].

The anomalous peak close to 110Oe is clearly seen for all temperatures. This feature is similar to previous observations [19]. The position of the peak is essentially temperature independent except for very close to $T_c$ when it shifts slightly to higher fields. However the simultaneous broadening of the peak introduces significant uncertainty in defining the actual peak position. As can be seen from Fig.4 this peak can be also detected in resistive measurements. The peak position is slightly higher and shifts to ~140Oe. The resistive commensurate peaks are very sharp. This is the case of dominant vortex-vortex interactions.

Further insight into these results can be obtained from the Cole-Cole plot [28] presented in Fig.5. We can see that this anomalous peak (indicated by arrow for each temperature) actually separates different regimes in the ac-shielding. The high field regions above the peak (Fig.3) merge to a universal curve close to theoretical results for a disc [30]. The low field oscilatory part follow another curve shifted to higher $\chi''$.

The ac-susceptibility of homogeneous superconductors has been extensively analysed theoretically [29, 30]. In the framework of Bean model the real and imaginary susceptibilities are expressed in terms of a dimensionless parameter $h = H_{ac} / H_p$. Here $H_p$ is the penetration field, which is proportional to the critical current. This corresponds to a universal curve parametrically determined by h and independent from critical current variations with magnetic field and temperature. This universal curve is very robust and shows only small changes for significant variations in topology of the sample or current-voltage characteristics [31]. Significant deviation from such universal behaviour, which we observe in Fig.6 suggests large changes either in the mechanism of the dissipation or shielding. We see two possible mechanisms involved. Appearance of interstitial vortices should produce increased dissipation and correspondingly smaller $\chi''$. Alternatively this anomaly may b e related to some changes in topology of shielding, which suppresses flux pentration into holes for fields below the peak.

## 4. Conclusions

We have studied the interactions of the vortex lattice with a periodic square array of holes in a superconducting Nb film using magnetic and resistive measurements. Three different temperature regions have been observed. Flux jumps dominate the low temperature region. Step-like commensurate anomalies are prominent at intermediate temperatures. Finally, sharp peaks develop approaching $T_c$. These transformations are governed by the interplay between vortex-vortex interactions and pinning by holes. In all measurements a peak has been observed close to commensurate states 5-7. The Cole-Cole plot reveals significant changes in the flux penetration mechanism at this point.

This work has been supported by the EPSRC (GR/N29396/01) and by the US NSF (DMR-0210519).

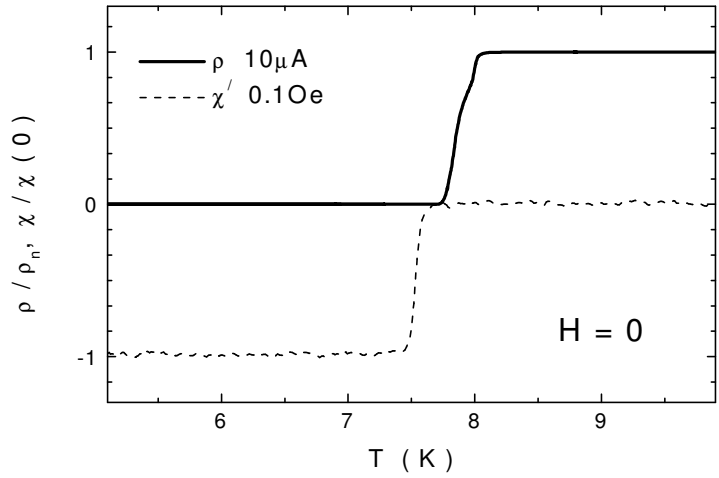

Fig.1 Superconducting transition from simultaneous measurements of resistivity and ac-susceptibility

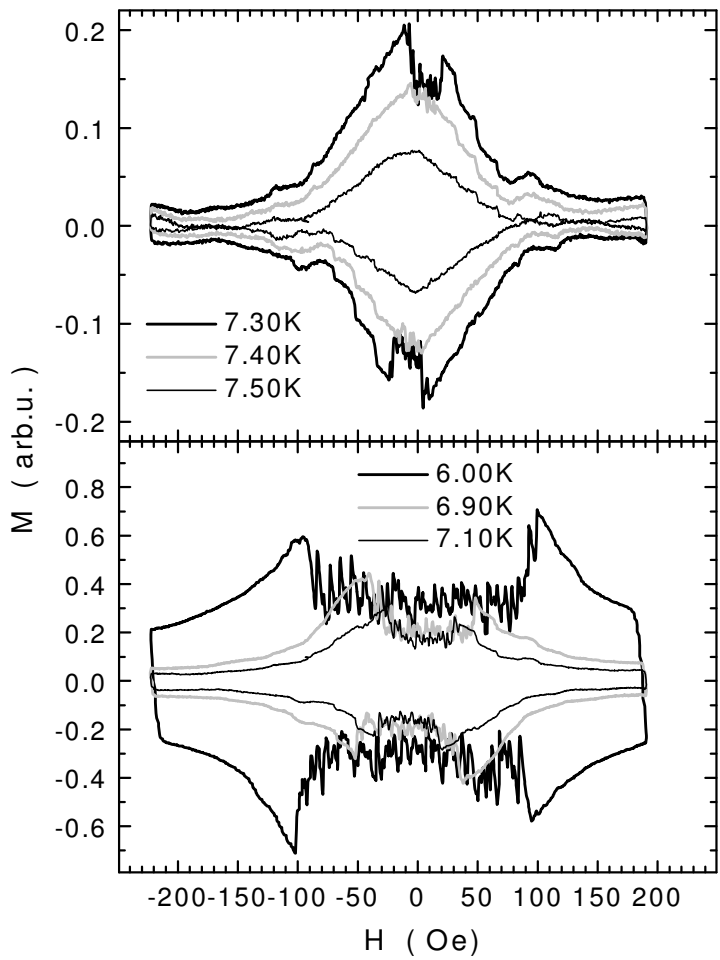

Fig.2 Magnetization curves at different temperatures

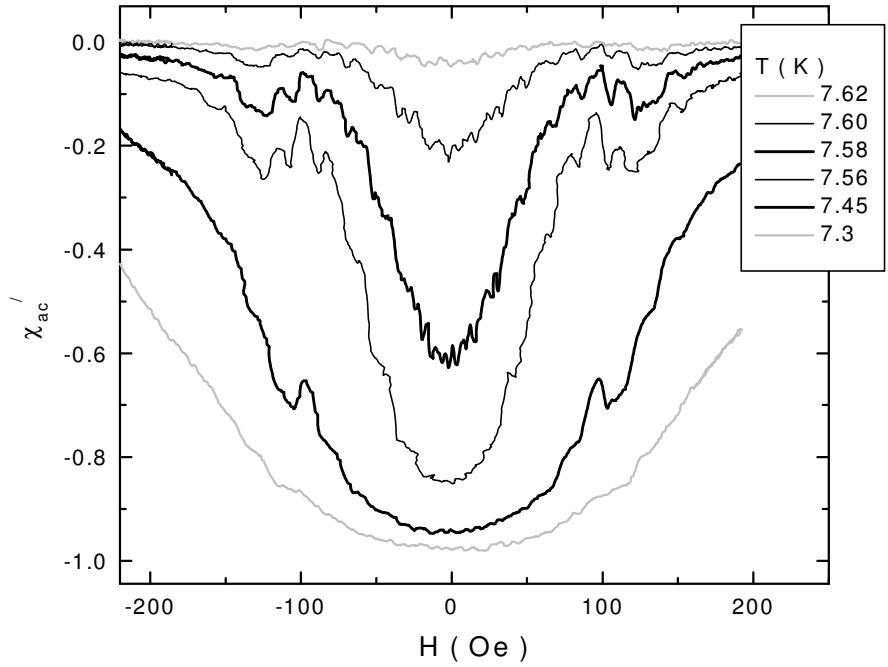

Fig.3 ac-susceptibility at different temperatures, $H_{ac}$= 0.24Oe.

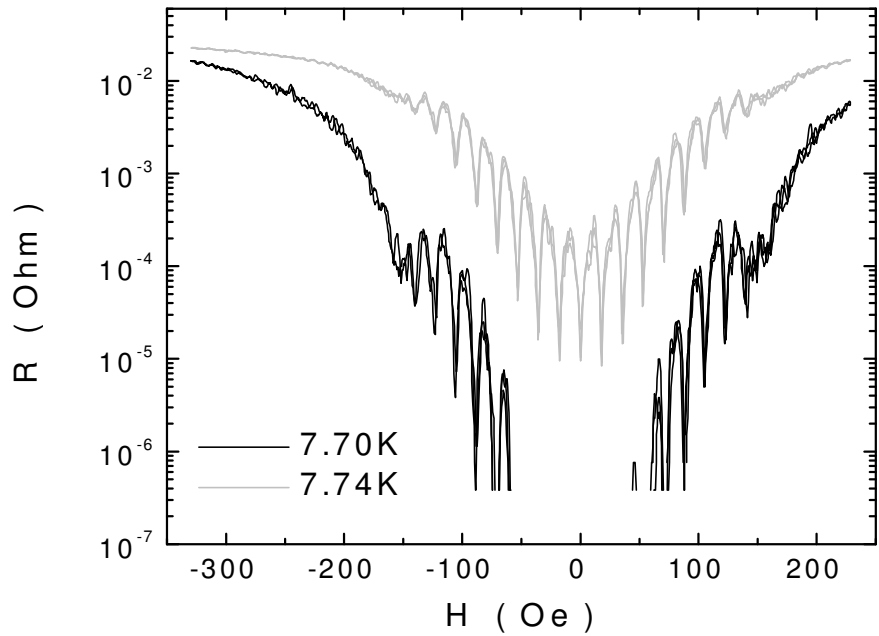

Fig.4 Magnetic field dependences of resistivity at different temperatures

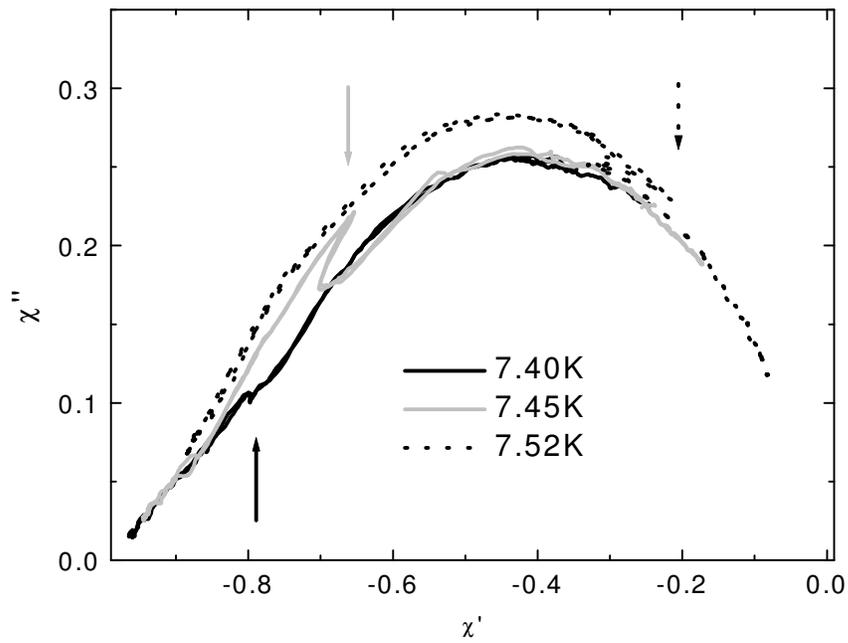

Fig.5 Cole-Cole plot for different temperatures. Arrows show positions of the peak for the corresponding temperatures.